\documentclass
[prd,amsfonts,amssymb,twocolumn,a4paper,balancelastpage,twoside]{revtex4}%
\usepackage{amsfonts}
\usepackage{amsmath}
\usepackage{amssymb}
\usepackage{graphicx}%
\providecommand{\U}[1]{\protect\rule{.1in}{.1in}}
\newtheorem{theorem}{Theorem}

\newtheorem{remark}[theorem]{Remark}

\begin{document}
\preprint{ }
\title[Bianchi II with time varying constants]{Bianchi II with time varying constants. Self-similar approach.}
\author{Jos\'e Antonio Belinch\'on}
\affiliation{Dept. F\'isica. ETS Arquitectura. UPM. Av. Juan de Herrera $4$. Madrid
$28040$. Espa\~{n}a}
\pacs{PACS number}

\begin{abstract}
We study a perfect fluid Bianchi II models with time varying constants under
the self-similarity approach. In the first of the studied model, we consider
that only vary $G$ and $\Lambda.$ The obtained solution is more general that
the obtained one for the classical solution since it is valid for an equation
of state $\omega\in\left(  -1,\infty\right)  $ while in the classical solution
$\omega\in\left(  -1/3,1\right)  .$ Taking into account the current
observations, we conclude that $G$ must be a growing time function while
$\Lambda$ is a positive decreasing function. In the second of the studied
models we consider a variable speed of light (VSL). We obtain a similar
solution as in the first model arriving to the conclusions that $c$ must be a
growing time function if $\Lambda$ is a positive decreasing function.
\end{abstract}
\date{\today}
\maketitle

\section{Introduction.}

Since the pioneering work of Dirac (\cite{1}), who proposed, motivated by the
occurrence of large numbers in Universe, a theory with a time variable
gravitational coupling constant $G$, cosmological models with variable $G$ and
nonvanishing and variable cosmological term, $\Lambda,$ have been intensively
investigated in the physical literature (see for example \cite{2}-\cite{14}).

In modern cosmological theories, the cosmological constant remains a focal
point of interest (see \cite{cc1}-\cite{cc4}\ \ for reviews of the problem). A
wide range of observations now compellingly suggest that the universe
possesses a non-zero cosmological constant. Some of the recent discussions on
the cosmological constant \textquotedblleft problem\textquotedblright\ and on
cosmology with a time-varying cosmological constant point out that in the
absence of any interaction with matter or radiation, the cosmological constant
remains a \textquotedblleft constant\textquotedblright. However, in the
presence of interactions with matter or radiation, a solution of Einstein
equations and the assumed equation of covariant conservation of stress-energy
with a time-varying $\Lambda$ can be found. This entails that energy has to be
conserved by a decrease in the energy density of the vacuum component followed
by a corresponding increase in the energy density of matter or radiation
\ Recent observations strongly favour a significant and a positive value of
$\Lambda$ with magnitude $\Lambda(G\hbar/c^{3})\approx10^{-123}$. These
observations suggest on accelerating expansion of the universe, $q<0$.

Our current understanding of the physical universe is anchored on the analysis
of expanding, isotropic and homogeneous models with a cosmological constant,
and linear perturbations thereof. This model successfully accounts for the
late time universe, as is evidenced by the observation of large scale cosmic
microwave background observations. Parameter determination from the analysis
of CMB fluctuations appears to confirm this picture. However, further analyses
seem to suggests some inconsistency. In particular, its appears that the
universe could have a preferred direction. Followup analyses of various sets
of WMAP data sets, with different techniques seem to lead to the same
conclusion. It is still unclear whether or not the directional preference is
intrinsic to the underlying model, and what implications this has on our
understanding of cosmology. For this reason Bianchi models are important in
the study of anisotropies.

Recently , the cosmological implications of a variable speed of light during
the early evolution of the Universe have been considered. Varying speed of
light (VSL) models proposed by Moffat (\cite{moff}) and Albrecht and Magueijo
(\cite{Magueijo0}), in which light was travelling faster in the early periods
of the existence of the Universe, might solve the same problems as inflation.
Hence they could become a valuable alternative explanation of the dynamics and
evolution of our Universe and provide an explanation for the problem of the
variation of the physical \textquotedblleft constants\textquotedblright.
Einstein's field equations (EFE) for FRW spacetime in the VSL theory have been
solved by Barrow (\cite{Barrow-c} and \cite{barrow b1}\ for anisotropic
models), who also obtained the rate of variation of the speed of light
required to solve the flatness and cosmological constant problems (see J.
Magueijo (\cite{Magueijo1}) for a review of these theories).

This model is formulated under the strong assumption that a $c$ variable
(where $c$ stands for the speed of light) does not introduce any corrections
into the curvature tensor, furthermore, such formulation does not verify the
covariance and the Lorentz invariance as well as the resulting field equations
do not verify the Bianchi identities either (see Bassett et al \cite{Bassett}).

Nevertheless, some authors (T. Harko and M. K. Mak \cite{Harko1}, P.P. Avelino
and C.J.A.P. Martins \cite{Avelino} and H. Shojaie et al \cite{Iranies}) have
proposed a new generalization of General Relativity which also allows
arbitrary changes in the speed of light, $c$, and the gravitational constant,
$G$, but in such a way that variations in the speed of light introduces
corrections to the curvature tensor in the Einstein equations in the
cosmological frame. This new formulation is both covariant and Lorentz
invariant and as we will show the resulting field equations (FE) verify the
Bianchi identities.

The purpose of this paper is to study a perfect fluid Bianchi II\ model with
time varying constants through the self-similarity (SS) hypothesis (see foe
example \cite{CT}-\cite{ColeyDS}). The study of SS models is quite important
since a large class of orthogonal spatially homogeneous models are
asymptotically self-similar at the initial singularity and are approximated by
exact perfect fluid or vacuum self-similar power law models. Exact
self-similar power-law models can also approximate general Bianchi models at
intermediate stages of their evolution. This last point is of particular
importance in relating Bianchi models to the real Universe. At the same time,
self-similar solutions can describe the behaviour of Bianchi models at late
times i.e. as $t\rightarrow\infty.$

We would like to emphasize that in this work we are more interesting in
mathematical respects (exact solutions) than in studying strong physical
consequences. Nevertheless we consider some observational data in order to
rule out some the obtained solutions.

Therefore the paper is organized as follows. In the second section, we begin
by defining the metric and calculating its Killing vectors as well as their
algebra. We also calculate the four velocity and the quantities derived from
it, i.e. the expansion, Hubble parameter etc... In section three, we review
some basic ideas about self-similarity. Once we have clarified some concepts
then we calculate the homothetic vector field as well as the constrains for
the scale factors. In section fourth we review the classical solution i.e. a
perfect fluid model with \textquotedblleft constant\textquotedblright%
\ constants. We show that the obtained solution belongs to the LRS\ Bianchi
type II by calculating the fourth Killing vector field. We also calculate all
the curvature invariants showing that the obtained solution is singular. These
results will be valid for the rest of the studied models. We end this section
showing that there is not solution for the vacuum model. In section five, we
study a perfect fluid model with variable $G$ and $\Lambda.$ We compare the
solution with the obtained one in the above section showing that we have been
able to relax the constrains obtained for the classical model. We rule out
some of the obtained solution taking into account the current observations
which suggest us that $\Lambda_{0}>0$ and $q<0.$ In section six, we study a
perfect fluid model with variable speed of light (VSL). We out line the field
equations taking into account the effects of a c-var into the curvature
tensor. This point is essential in this approach as we have showed in a
previous paper (\cite{tony2}). We calculate the homothetic vector field for
the new metric and we show the new constrains for the scale factors. If we
assume a particular solution for $c(t)$ then we show that it may be a
decreasing time function. Only taking into account observational restriction
we are able to rule out some of the obtained solutions as in the above
section. We end summarizing the conclusions in the last section.

\section{The metric.}

Throughout the paper $M$ will denote the usual smooth (connected, Hausdorff,
4-dimensional) spacetime manifold with smooth Lorentz metric $g$ of signature
$(-,+,+,+)$ (see for example \cite{MC}). Thus $M$ is paracompact. A comma,
semi-colon and the symbol $\mathcal{L}$ denote the usual partial, covariant
and Lie derivative, respectively, the covariant derivative being with respect
to the Levi-Civita connection on $M$ derived from $g$. The associated Ricci
and stress-energy tensors will be denoted in component form by $R_{ij}(\equiv
R^{c}{}_{jcd})$ and $T_{ij}$ respectively.

A Bianchi II space-time is a spatially homogeneous space-time which admits an
abelian group of isometries $G_{3}$, acting on spacelike hypersurfaces,
generated by the spacelike KVs,%

\begin{equation}
\xi_{1}=K\partial_{x},\qquad\xi_{2}=\partial_{y},\qquad\xi_{3}=-Ky\partial
_{x}+\partial_{z}, \label{killings}%
\end{equation}
and their algebra is:%
\[%
\begin{array}
[c]{c|ccc}
& \xi_{1} & \xi_{2} & \xi_{3}\\\hline
\xi_{1} & 0 & 0 & 0\\
\xi_{2} & 0 & 0 & \xi_{1}\\
\xi_{3} & 0 & -\xi_{1} & 0
\end{array}
\qquad\left[  \xi_{i},\xi_{j}\right]  =C_{ij}^{k}\xi_{k},\qquad C_{23}^{1}=1.
\]

In synchronous co-ordinates the metric is:%
\begin{align}
ds^{2}  &  =-c^{2}dt^{2}+a^{2}(t)dx^{2}+\left(  b^{2}(t)+K^{2}z^{2}%
a^{2}(t)\right)  dy^{2}+\nonumber\\
&  +2Ka^{2}(t)zdxdy+d^{2}(t)dz^{2} \label{metric}%
\end{align}
where the metric functions $a(t),b(t),d(t)$ are functions of the time
co-ordinate only and $K\in\mathbb{R}$. The introduction of this constant is
essential since if we set $K=1$, as it is the usual way, then there is not SS
solution for the outlined field equations. In this paper we are interested
only in Bianchi II space-times, hence all metric functions are assumed to be
different and the dimension of the group of isometries acting on the spacelike
hypersurfaces is three.

Once we have defined the metric and we know which are its killing vectors,
then we calculate the four velocity. It must verify, $\mathcal{L}_{\xi_{i}%
}u_{i}=0,$ so we may define the four velocity as follows:%
\begin{equation}
u^{i}=\left(  \frac{1}{c},0,0,0\right)  ,
\end{equation}
in such a way that it is verified, $g(u^{i},u^{i})=-1.$

From the definition of the 4-velocity we find that:
\begin{align*}
\theta &  =u_{\,\,;i}^{i}=\frac{1}{c}\left(  \frac{a^{\prime}}{a}%
+\frac{b^{\prime}}{b}+\frac{d^{\prime}}{d}\right)  =\frac{1}{c}H,\\
H  &  =H_{1}+H_{2}+H_{3},\\
q  &  =\frac{d}{dt}\left(  \frac{1}{H}\right)  -1,\\
\sigma^{2}  &  =\frac{1}{3c^{2}}\left(  H_{1}^{2}+H_{2}^{2}+H_{3}^{2}%
-H_{1}H_{2}-H_{1}H_{3}-H_{2}H_{3}\right)  .
\end{align*}

\section{Self-similarity.}

In general relativity, the term self-similarity can be used in two ways. One
is for the properties of space-times, the other is for the properties of
matter fields. These are not equivalent in general. The self-similarity in
general relativity was defined for the first time by Cahill and Taub (see
\cite{CT}, and for general reviews \cite{21}-\cite{Hall}). Self-similarity is
defined by the existence of a homothetic vector ${V}$ in the spacetime, which
satisfies
\begin{equation}
\mathcal{L}_{V}g_{ij}=2\alpha g_{ij}, \label{gss1}%
\end{equation}
where $g_{ij}$ is the metric tensor, $\mathcal{L}_{V}$ denotes Lie
differentiation along ${V}$ and $\alpha$ is a constant. This is a special type
of conformal Killing vectors. This self-similarity is called homothety. If
$\alpha\neq0$, then it can be set to be unity by a constant rescaling of ${V}%
$. If $\alpha=0$, i.e. $\mathcal{L}_{V}g_{ij}=0$, then ${V}$ is a Killing vector.

Homothety is a purely geometric property of spacetime so that the physical
quantity does not necessarily exhibit self-similarity such as $\mathcal{L}%
_{V}Z=dZ$, where $d$ is a constant and $Z$ is, for example, the pressure, the
energy density and so on. From equation (\ref{gss1}) it follows that
$\mathcal{L}_{V}R^{i}\,_{jkl}=0,$ and hence $\mathcal{L}_{V}R_{ij}=0,$and
$\mathcal{L}_{V}G_{ij}=0.$ A vector field ${V}$ that satisfies the above
equations is called a curvature collineation, a Ricci collineation and a
matter collineation, respectively. It is noted that such equations do not
necessarily mean that ${V}$ is a homothetic vector. We consider the Einstein
equations $G_{ij}=8\pi GT_{ij},$ where $T_{ij}$ is the energy-momentum tensor.
If the spacetime is homothetic, the energy-momentum tensor of the matter
fields must satisfy $\mathcal{L}_{V}T_{ij}=0$. For a perfect fluid case, the
energy-momentum tensor takes the form $T_{ij}=(p+\rho)u_{i}u_{j}+pg_{ij}%
,$where $p$ and $\rho$ are the pressure and the energy density, respectively.
Then, equations~(\ref{gss1}) result in
\begin{equation}
\mathcal{L}_{V}u^{i}=-\alpha u^{i},\quad\mathcal{L}_{V}\rho=-2\alpha\rho
,\quad\mathcal{L}_{V}p=-2\alpha p. \label{ssmu}%
\end{equation}

As shown above, for a perfect fluid, the self-similarity of the spacetime and
that of the physical quantity coincide. However, this fact does not
necessarily hold for more general matter fields. Thus the self-similar
variables can be determined from dimensional considerations in the case of
homothety. Therefore, we can conclude homothety as the general relativistic
analogue of complete similarity.

From the constraints (\ref{ssmu}), we can show that if we consider the
barotropic equation of state, i.e., $p=f(\rho)$, then the equation of state
must have the form $p=\omega\rho$, where $\omega$ is a constant. In this
paper, we would like to try to show how taking into account this class of
hypothesis one is able to find exact solutions to the field equations within
the framework of the time varying constants.

From equation (\ref{gss1})$,$we find the following homothetic vector field for
the BII metric (\ref{metric}):%
\begin{equation}
V=t\partial_{t}+\left(  1-t\frac{a^{\prime}}{a}\right)  x\partial_{x}+\left(
1-t\frac{b^{\prime}}{b}\right)  y\partial_{y}+\left(  1-t\frac{d^{\prime}}%
{d}\right)  z\partial_{z}, \label{HO}%
\end{equation}
with the following constrains for the scale factors:
\begin{equation}
a(t)=a_{0}t^{a_{1}},\qquad b(t)=b_{0}t^{a_{2}},\qquad d(t)=d_{0}t^{a_{3}},
\label{scales}%
\end{equation}
with $a_{1},a_{2},a_{3}\in\mathbb{R},$ in such a way, that the constants
$a_{i}$ must verify the following restriction:%
\begin{equation}
a_{2}+a_{3}-a_{1}=1. \label{restriction}%
\end{equation}
therefore the resulting homothetic vector field is:%
\begin{equation}
V=t\partial_{t}+\left(  1-a_{1}\right)  x\partial_{x}+\left(  1-a_{2}\right)
y\partial_{y}+\left(  1-a_{3}\right)  z\partial_{z},
\end{equation}
which is the same as the found one in the Bianchi I model (see for example
\cite{tony1}).

\section{The perfect fluid Model. The classical solution.}

Taking into account the field equations (FE)%
\begin{equation}
R_{ij}-\frac{1}{2}Rg_{ij}=\frac{8\pi G}{c^{4}}T_{ij}-\Lambda g_{ij},
\end{equation}
where, in this section, we shall consider that $\Lambda$ vanish, and the
energy-momentum tensor, $T_{ij},$ is defined as follows:%
\begin{equation}
T_{ij}=(\rho+p)u_{i}u_{j}+pg_{ij}, \label{eq11}%
\end{equation}
with the usual equation of state $p=\omega\rho,$ $\omega\in\mathbb{R}$, as we
have discussed above$.$

The resulting FE for the metric (\ref{metric}) with a perfect fluid matter
model (\ref{eq11}) are:%

\begin{align}
\frac{a^{\prime}}{a}\frac{b^{\prime}}{b}+\frac{a^{\prime}}{a}\frac{d^{\prime}%
}{d}+\frac{d^{\prime}}{d}\frac{b^{\prime}}{b}-\frac{K^{2}}{4}\frac{a^{2}c^{2}%
}{b^{2}d^{2}}  &  =\frac{8\pi G}{c^{2}}\rho,\label{p1}\\
\frac{b^{\prime\prime}}{b}+\frac{d^{\prime\prime}}{d}+\frac{d^{\prime}}%
{d}\frac{b^{\prime}}{b}-\frac{3K^{2}}{4}\frac{a^{2}c^{2}}{b^{2}d^{2}}  &
=-\frac{8\pi G}{c^{2}}\omega\rho,\label{p2}\\
\frac{d^{\prime\prime}}{d}+\frac{a^{\prime\prime}}{a}+\frac{a^{\prime}}%
{a}\frac{d^{\prime}}{d}+\frac{K^{2}}{4}\frac{a^{2}c^{2}}{b^{2}d^{2}}  &
=-\frac{8\pi G}{c^{2}}\omega\rho,\label{p3}\\
\frac{b^{\prime\prime}}{b}+\frac{a^{\prime\prime}}{a}+\frac{a^{\prime}}%
{a}\frac{b^{\prime}}{b}+\frac{K^{2}}{4}\frac{a^{2}c^{2}}{b^{2}d^{2}}  &
=-\frac{8\pi G}{c^{2}}\omega\rho,\label{p4}\\
\rho^{\prime}+\rho\left(  1+\omega\right)  \left(  \frac{a^{\prime}}{a}%
+\frac{b^{\prime}}{b}+\frac{d^{\prime}}{d}\right)   &  =0. \label{p5}%
\end{align}

Now, if we take into account the obtained SS restrictions for the scale
factors i.e.
\[
a(t)=a_{0}t^{a_{1}},\qquad b(t)=b_{0}t^{a_{2}},\qquad d(t)=d_{0}t^{a_{3}},
\]
where $a_{1},a_{2},a_{3}\in\mathbb{R},$\ must satisfy $a_{2}+a_{3}-a_{1}=1,$
we get from eq. (\ref{p5}):%
\begin{equation}
\rho=\rho_{0}t^{-\gamma},
\end{equation}
where $\gamma=\alpha\left(  1+\omega\right)  =\left(  a_{1}+a_{2}%
+a_{3}\right)  \left(  1+\omega\right)  .$ Therefore the system to solve is
the following one:%
\begin{align}
a_{2}(a_{2}-1)+a_{3}\left(  a_{3}-1\right)  +a_{2}a_{3}-\frac{3K^{2}}{4}  &
=-A\omega,\\
a_{3}\left(  a_{3}-1\right)  +a_{1}a_{3}+a_{1}\left(  a_{1}-1\right)
+\frac{K^{2}}{4}  &  =-A\omega,\\
a_{2}(a_{2}-1)+a_{1}a_{2}+a_{1}(a_{1}-1)+\frac{K^{2}}{4}  &  =-A\omega,\\
a_{1}+a_{2}+a_{3}  &  =\frac{2}{1+\omega},
\end{align}
where we have set the constant $A$ as follows
\[
A=a_{1}a_{2}+a_{1}a_{3}+a_{3}a_{2}-\frac{K^{2}}{4},
\]
and $a_{2}+a_{3}-a_{1}=1.$

The obtained solution is the following one:%
\[
a_{1}=\frac{1-\omega}{2\left(  \omega+1\right)  },\qquad a_{2}=a_{3}%
=\frac{\omega+3}{4\left(  \omega+1\right)  },
\]
therefore \ the constrain $a_{2}+a_{3}-a_{1}=1$ collapses to $2a_{2}-a_{1}=1,$
and%
\[
K^{2}=\frac{1+2\omega-3\omega^{2}}{4\left(  \omega+1\right)  ^{2}}%
,\qquad\Longleftrightarrow\qquad\omega\in\left(  -\frac{1}{3},1\right)  ,
\]
so $a_{2}\in\left(  \frac{1}{2},1\right)  ,\quad a_{1}\in\left(  0,1\right)
,\quad K^{2}\in\left(  0,1/4\right)  ,$ and $A=2a_{1}a_{2}+a_{2}^{2}%
-\frac{K^{2}}{4}=\frac{\left(  5-\omega\right)  }{4\left(  \omega+1\right)
^{2}}.$ Taking into account all these results the homothetic vector field
collapses to%
\[
V=t\partial_{t}+\left(  1-a_{1}\right)  x\partial_{x}+\left(  1-a_{2}\right)
\left(  y\partial_{y}+z\partial_{z}\right)  ,
\]
this is the already solution obtained by Hsu et al (\cite{HW}). This solution
has been already obtained by Collins (\cite{Collins})

\begin{remark}
As is it observed, we may find the relation $a_{2}+a_{3}-a_{1}=1$ from
dimensional considerations, only looking at the FE. Note that the quantity%
\[
\frac{a^{2}c^{2}}{b^{2}d^{2}}\thickapprox\frac{a_{0}^{2}t^{2a_{1}}}{b_{0}%
^{2}t^{2a_{2}}d_{0}^{2}t^{2a_{3}}}\thickapprox t^{2a_{1}-2a_{2}-2a_{3}%
}\thickapprox t^{-2},
\]
after substitution, it must have dimensions of $T^{-2}$ as the rest of the
factors. Therefore we find the relation $a_{2}+a_{3}-a_{1}=1.$
\end{remark}

Therefore we have obtained a LRS\ BII model, note that $d=b,$ and hence the
metric collapses to this one:%
\begin{align}
ds^{2}  &  =-c^{2}dt^{2}+a^{2}(t)dx^{2}+\left(  b^{2}(t)+K^{2}z^{2}%
a^{2}(t)\right)  dy^{2}+\nonumber\\
&  +2Ka^{2}(t)zdxdy+b^{2}(t)dz^{2}, \label{LRS BII}%
\end{align}
which admits the three Killings vectors (\ref{killings}) and this new one%

\begin{equation}
\xi_{4}=-K\left(  \frac{z^{2}}{2}-\frac{y^{2}}{2}\right)  \partial
_{x}+z\partial_{y}-y\partial_{z}.
\end{equation}

Hence their algebra is:%
\[%
\begin{array}
[c]{c|cccc}
& \xi_{1} & \xi_{2} & \xi_{3} & \xi_{4}\\\hline
\xi_{1} & 0 & 0 & 0 & 0\\
\xi_{2} & 0 & 0 & \xi_{1} & -\xi_{3}\\
\xi_{3} & 0 & -\xi_{1} & 0 & \xi_{2}\\
\xi_{4} & 0 & \xi_{3} & -\xi_{2} & 0
\end{array}
\]
i.e. $C_{23}^{1}=1,\quad C_{24}^{3}=-1,$\ and $C_{34}^{2}=1.$

With the obtained results we can see that
\begin{align*}
H  &  =\frac{4a_{2}-1}{t}=\left(  \frac{2}{\omega+1}\right)  \frac{1}{t},\\
q  &  =-\frac{4a_{2}}{4a_{2}-1}=-\frac{3+\omega}{2}<0,\qquad\forall\omega
\in\left(  -1/3,1\right)  ,
\end{align*}
while the shear behaves as:%
\[
\sigma^{2}=\frac{\left(  a_{2}-1\right)  ^{2}}{3c^{2}t^{2}}=\left(
\frac{3\omega+1}{4\left(  \omega+1\right)  }\right)  ^{2}\frac{1}{3c^{2}t^{2}%
}.
\]
As it is observed, these quantities fit perfectly with the current
observations by High -Z Supernova Team and Supernova Cosmological Project see
for example: Garnavich \textit{et al.}, 1998 \cite{G}; Perlmutter \textit{et
al.}, 1997, 1998, 1999 \cite{Per}; Riess \textit{et al.}, 1998 \cite{Riess};
Schmidt \textit{et al.}, 1998 \cite{Sch}.

\subsection{Curvature behaviour.}

With all these results, we find the following behaviour for the curvature
invariants (see for example \cite{Caminati}-\cite{Barrow}).

Ricc Scalar, $I_{0},$ yields%
\begin{equation}
I_{0}=\frac{2}{c^{2}t^{2}}\left(  11a_{2}^{2}-10a_{2}+2-\frac{K^{2}c^{2}}%
{4}\right)  ,
\end{equation}
while Krestchmann scalar, $I_{1}:=R_{ijkl}R^{ijkl}$, yields:%
\begin{align*}
I_{1}  &  =\frac{1}{4c^{4}t^{4}}\left(  432a_{2}^{4}-960a_{2}^{3}+896a_{2}%
^{2}-384a_{2}+64\right. \\
&  \left.  +K^{2}c^{2}\left(  11K^{2}c^{2}-40a_{2}^{2}+80a_{2}-48\right)
\right)  .
\end{align*}

The full contraction of the Ricci tensor, $I_{2}:=R_{ij}R^{ij},$ is:%
\begin{align*}
I_{2}  &  =\frac{1}{4c^{4}t^{4}}\left(  528a_{2}^{4}-1024a_{2}^{3}%
+768a_{2}^{2}-256a_{2}+32\right. \\
&  \left.  K^{2}c^{2}\left(  3K^{2}c^{2}-16a_{2}+8\right)  \right)  ,
\end{align*}
this means that the model is singular.

The non-zero components of the Weyl tensor. The following components of the
Weyl tensor run to $\pm\infty$ when $t\rightarrow0,$%
\[
C_{txtx}\thickapprox C_{txty}\thickapprox t^{4\left(  a_{2}-1\right)  },\quad
C_{tztz}\thickapprox t^{2\left(  a_{2}-1\right)  },
\]
and
\[
C_{tyty}\thickapprox t^{2\left(  a_{2}-1\right)  }+z^{2}t^{4\left(
a_{2}-1\right)  },
\]
these others depend on the value of $a_{2},$%
\[
C_{txyz}\thickapprox C_{tyxz}\thickapprox C_{tyyz}\thickapprox C_{tzxy}%
\thickapprox t^{4a_{2}-3},
\]%
\[
C_{xzyz}\thickapprox C_{xzxz}\thickapprox C_{xyxy}\thickapprox t^{2\left(
3a_{2}-2\right)  },
\]%
\[
C_{yzyz}\thickapprox t^{2\left(  2a_{2}-1\right)  }+z^{2}t^{2\left(
3a_{2}-2\right)  }.
\]
i.e. they may run to zero as well as to $\pm\infty$ when $t\rightarrow0,$
depending on the value of $a_{2}.$

The Weyl scalar, $I_{3}=C^{abcd}C_{abcd}=I_{1}-2I_{2}+\frac{1}{3}I_{0}^{2},$
(this definition is only valid when $n=4)$%
\begin{align*}
I_{3}  &  =\frac{4}{3c^{4}t^{4}}\left(  \left[  \left(  a_{2}-1\right)
\left(  -2\left(  a_{2}-1\right)  -3Kc\right)  +K^{2}c^{2}\right]  \right. \\
&  \left.  \times\left[  \left(  a_{2}-1\right)  \left(  -2\left(
a_{2}-1\right)  +3Kc\right)  +K^{2}c^{2}\right]  \right)  .
\end{align*}

The electric part scalar $I_{4}=E_{ij}E^{ij},$ (see W.C. Lim et al
\cite{Coley})%
\[
I_{4}=\frac{1}{6c^{4}t^{4}}\left(  -2\left(  a_{2}-1\right)  ^{2}+K^{2}%
c^{2}\right)  ^{2},
\]
while the magnetic part scalar $I_{5}=H_{ij}H^{ij},$ yields%
\[
I_{5}=\frac{3}{2c^{2}t^{4}}\left(  a_{2}-1\right)  ^{2}K^{2}.
\]

Therefore the Weyl parameter $W^{2}$ (see W.C. Lim et al \cite{Coley})%
\[
W^{2}=\frac{1}{6}\left(  E_{ij}E^{ij}+H_{ij}H^{ij}\right)  ,
\]
therefore%
\[
W^{2}=\frac{1}{36c^{4}t^{4}}\left(  \left(  a_{2}-1\right)  ^{2}+K^{2}%
c^{2}\right)  ^{2}\left(  \left(  2a_{2}-2\right)  ^{2}+K^{2}c^{2}\right)
^{2},
\]
note that the value of $W^{2}$ is really small.

The gravitational entropy (see \cite{gron1}-\cite{gron2})
\[
P^{2}=\frac{I_{3}}{I_{2}}=\frac{I_{1}-2I_{2}+\frac{1}{3}I_{0}^{2}}{I_{2}%
}=\frac{I_{1}}{I_{2}}+\frac{1}{3}\frac{I_{0}^{2}}{I_{2}}-2=const.,
\]
since the spacetime is SS (see \cite{lake} and \cite{tony1} for a discussion).

\subsection{Vacuum model.}

In this section we shall show that there is no self-similar solution for the
vacuum model. In this case, after substitution, the resulting FE are:%

\begin{align}
a_{1}a_{2}+a_{1}a_{3}+a_{3}a_{2}-\frac{K^{2}}{4}  &  =0,\\
a_{2}(a_{2}-1)+a_{3}\left(  a_{3}-1\right)  +a_{2}a_{3}-\frac{3K^{2}}{4}  &
=0,\\
a_{3}\left(  a_{3}-1\right)  +a_{1}a_{3}+a_{1}\left(  a_{1}-1\right)
+\frac{K^{2}}{4}  &  =0,\\
a_{2}(a_{2}-1)+a_{1}a_{2}+a_{1}(a_{1}-1)+\frac{K^{2}}{4}  &  =0,
\end{align}
so, by solving this system, we only obtain the following unphysical solutions:%
\begin{align*}
a_{1}  &  =a_{2}=a_{3}=K=0,\\
a_{1}  &  =1,\qquad a_{2}=a_{3}=K=0,\\
a_{1}  &  =-\frac{1}{3},\qquad a_{2}=a_{3}=\frac{1}{3},\qquad K^{2}=-\frac
{4}{9},\\
a_{1}  &  =-\frac{1}{3},\qquad a_{2}=a_{3}=\frac{1}{3},\qquad K=0.
\end{align*}

Therefore, we conclude that there is no physical solution for the vacuum
model. Note that when we set $K=0,$ the metric is reduced to Bianchi I one. In
the same way, it is also verified the relation $a_{2}+a_{3}-a_{1}=1.$

\section{Perfect fluid model with G and Lambda variable.}

The resulting FE for the metric (\ref{metric}) with a perfect fluid matter
model (\ref{eq11}) and with $G$ and $\Lambda$ time-varying are:%

\begin{align}
\frac{a^{\prime}}{a}\frac{b^{\prime}}{b}+\frac{a^{\prime}}{a}\frac{d^{\prime}%
}{d}+\frac{d^{\prime}}{d}\frac{b^{\prime}}{b}-\frac{K^{2}}{4}\frac{a^{2}c^{2}%
}{b^{2}d^{2}}  &  =\frac{8\pi G}{c^{2}}\rho+\Lambda c^{2},\label{t1}\\
\frac{b^{\prime\prime}}{b}+\frac{d^{\prime\prime}}{d}+\frac{d^{\prime}}%
{d}\frac{b^{\prime}}{b}-\frac{3K^{2}}{4}\frac{a^{2}c^{2}}{b^{2}d^{2}}  &
=-\frac{8\pi G}{c^{2}}\omega\rho+\Lambda c^{2},\label{t2}\\
\frac{d^{\prime\prime}}{d}+\frac{a^{\prime\prime}}{a}+\frac{a^{\prime}}%
{a}\frac{d^{\prime}}{d}+\frac{K^{2}}{4}\frac{a^{2}c^{2}}{b^{2}d^{2}}  &
=-\frac{8\pi G}{c^{2}}\omega\rho+\Lambda c^{2},\label{t3}\\
\frac{b^{\prime\prime}}{b}+\frac{a^{\prime\prime}}{a}+\frac{a^{\prime}}%
{a}\frac{b^{\prime}}{b}+\frac{K^{2}}{4}\frac{a^{2}c^{2}}{b^{2}d^{2}}  &
=-\frac{8\pi G}{c^{2}}\omega\rho+\Lambda c^{2},\label{t4}\\
\rho^{\prime}+\rho\left(  1+\omega\right)  \left(  \frac{a^{\prime}}{a}%
+\frac{b^{\prime}}{b}+\frac{d^{\prime}}{d}\right)   &  =0,\label{t5}\\
\Lambda^{\prime}  &  =-\frac{8\pi}{c^{4}}G^{\prime}\rho. \label{t6}%
\end{align}

Now, we shall take into account the obtained SS restrictions for the scale
factors i.e.
\[
a(t)=a_{0}t^{a_{1}},\qquad b(t)=b_{0}t^{a_{2}},\qquad d(t)=d_{0}t^{a_{3}},
\]
where $a_{1},a_{2},a_{3}\in\mathbb{R},$ in such a way that they must verify
$a_{2}+a_{3}-a_{1}=1.$

From eq. (\ref{t5}) we get%
\begin{equation}
\rho=\rho_{0}t^{-\gamma}, \label{t_den}%
\end{equation}
where $\gamma=\left(  \omega+1\right)  \alpha$ and $\alpha=\left(  a_{1}%
+a_{2}+a_{3}\right)  $.

From eq. (\ref{t1}) we obtain:%
\begin{equation}
\Lambda=\frac{1}{c^{2}}\left[  At^{-2}-\frac{8\pi G}{c^{2}}\rho_{0}t^{-\left(
\omega+1\right)  \alpha}\right]  \label{peta1}%
\end{equation}
where $A=a_{1}a_{2}+a_{1}a_{3}+a_{2}a_{3}-\frac{K^{2}}{4}.$

Now, taking into account eq. (\ref{t6}) and eq. (\ref{peta1}), algebra brings
us to obtain%
\begin{equation}
G=G_{0}t^{\gamma-2}, \label{G}%
\end{equation}
where $G_{0}=\frac{c^{2}A}{4\pi\rho_{0}\left(  \omega+1\right)  \alpha}.$

While the cosmological \textquotedblleft constant\textquotedblright\ behaves
as:
\begin{equation}
\Lambda=\frac{A}{c^{2}}\left(  1-\frac{2}{\left(  \omega+1\right)  \alpha
}\right)  t^{-2}=\Lambda_{0}t^{-2}.
\end{equation}

With all these result we find that the system to solve is the following one:%
\begin{align}
a_{2}(a_{2}-1)+a_{3}\left(  a_{3}-1\right)  +a_{2}a_{3}-\frac{3K^{2}}{4}  &
=A\left(  \frac{\alpha-2}{\alpha}\right)  ,\\
a_{3}\left(  a_{3}-1\right)  +a_{1}a_{3}+a_{1}\left(  a_{1}-1\right)
+\frac{K^{2}}{4}  &  =A\left(  \frac{\alpha-2}{\alpha}\right)  ,\\
a_{2}(a_{2}-1)+a_{1}a_{2}+a_{1}(a_{1}-1)+\frac{K^{2}}{4}  &  =A\left(
\frac{\alpha-2}{\alpha}\right)  ,
\end{align}
whose solution is:%
\begin{equation}
a_{1}=2a_{2}-1,\quad a_{2}=a_{3},\quad K^{2}=-4a_{2}^{2}+6a_{2}-2,
\label{sol GL}%
\end{equation}
therefore this solution has only sense if $a_{2}\in\left(  \frac{1}%
{2},1\right)  .$ Note that for these values, $a_{1}>0,$ $a_{1}=2a_{2}%
-1\in\left(  0,1\right)  .$ This result is valid for all equation of state. We
would like to point out that this solution is more general that the obtained
one in the perfect fluid case with \textquotedblleft
constant\textquotedblright\ constants, since it is valid for $\omega\in\left(
-1,\infty\right)  ,$ although we are only interested in $\omega\in\left(
-1,a\right]  ,$ $a\geq1.$ It is important to emphasize that $\omega\neq-1.$

With this solution we may see that all the quantities depend on two variables,
$a_{2}$ and $\omega.$ Newton gravitational constant behaves as follows:%
\begin{equation}
G=\frac{c^{2}A}{4\pi\rho_{0}\left(  \omega+1\right)  \alpha}t^{\gamma-2}%
=G_{0}t^{g},\qquad G_{0}>0,
\end{equation}
with $g=\left(  \omega+1\right)  \left(  4a_{2}-1\right)  -2\in\left(
-2,4\right)  ,$ and $A=5a_{2}^{2}-2a_{2}-\frac{K^{2}}{4}=6a_{2}^{2}-\frac
{7}{2}a_{2}+\frac{1}{2}\in\left(  \frac{1}{4},3\right)  .$ So depending of the
different combinations between $a_{2}$ and $\omega,$ $G$ may be a growing or a
decreasing time function.

With regard to the behaviour of the cosmological constant, as we can see, it
is a decreasing time function, as it is expected. Only rest to know which is
the sign of $\Lambda_{0}$. We may express its behaviour in the following
tables%
\[%
\begin{array}
[c]{|c|c|c|}\hline
\omega & a_{2} & \Lambda_{0}\\\hline\hline
\left(  -1,1\right)  & \searrow1/2 & \leq0\\
1 & \searrow1/2 & 0\\
>1 & \searrow1/2 & \nearrow1\\\hline
\end{array}
\qquad%
\begin{array}
[c]{|c|c|c|}\hline
\omega & a_{2} & \Lambda_{0}\\\hline\hline
\left(  -1,-1/2\right)  & \nearrow1 & <0\\
-1/2 & \nearrow1 & 0\\
>-1/2 & \nearrow1 & \nearrow1\\\hline
\end{array}
\]
this means, for example, that when we fix $a_{2}\searrow1/2$ i.e. that $a_{2}$
tends to $1/2,$ for all value of $\omega$ between -1 and 1, $\Lambda_{0}<0,$
$\Lambda_{0}=0$, if $\omega=1$ and only $\Lambda_{0}>0$ when $\omega>1.$ In
the say way we may say that if we fix $a_{2}\nearrow1$, then $\forall\omega
\in\left(  -1,-1/2\right)  ,$ we find that $\Lambda_{0}<0.$ $\Lambda_{0}=0,$
if $\omega=-1/2$ and only it is found $\Lambda_{0}>0$ when $\omega>-1/2.$ So
may only say that the sign of $\Lambda_{0},$ depends on the parameters $a_{2}$
and $\omega.$ If we take into account the current observations which suggest
that $\Lambda_{0}>0,$ then we may rule out the values which make $\Lambda
_{0}<0.$

As it is observed the behaviour of $G$ and the sign of $\Lambda_{0}$, are
related. If $G$ is growing then $\Lambda_{0}>0,$ if $G=const.$ then
$\Lambda_{0}=0,$ i.e. the model collapses to the standard one studied above,
and if $G$ is decreasing then $\Lambda_{0}<0.$ Therefore if we consider the
recent observations which suggest us that $\Lambda_{0}>0,$ we must rule out
the other cases, concluding that $G$ is a growing time function and $\Lambda$
is a positive decreasing function.

The energy-density behaves as%
\begin{equation}
\rho=\rho_{o}t^{-\gamma},\qquad\text{with}\qquad\gamma\in\left(  0,6\right)
\end{equation}
so it is always a time decreasing function if $\omega>-1,$ and to end, we find
the following behaviour for the Hubble, deceleration and shear parameters:%

\[
H=\frac{4a_{2}-1}{t},\qquad q=-\frac{4a_{2}}{4a_{2}-1}<0,
\]%
\begin{equation}
\sigma^{2}=\frac{\left(  a_{2}-1\right)  ^{2}}{3c^{2}t^{2}}\longrightarrow0.
\end{equation}

As in the above model, the metric of this one collapses to a LRSBII type
(\ref{LRS BII}).

\section{Perfect fluid model with VSL.}

We start by defining the new metric as follows:%
\begin{align}
ds^{2}  &  =-c(t)^{2}dt^{2}+a^{2}(t)dx^{2}+\left(  b^{2}(t)+K^{2}z^{2}%
a^{2}(t)\right)  dy^{2}+\nonumber\\
&  +2Ka^{2}(t)zdxdy+d^{2}(t)dz^{2}, \label{LRSBII-cvar}%
\end{align}
note that we have simply replaced $c$ by $c(t),$ so we shall consider the four
velocity:%
\begin{equation}
u=\left(  \frac{1}{c(t)},0,0,0\right)  ,\quad/\qquad u_{i}u^{i}=-1.
\end{equation}

The time derivatives of $G,c$ and $\Lambda$ are related by the Bianchi
identities
\begin{equation}
\left(  R_{ij}-\frac{1}{2}Rg_{ij}\right)  ^{;j}=\left(  \frac{8\pi G}{c^{4}%
}T_{ij}-\Lambda g_{ij}\right)  ^{;j}, \label{eq8}%
\end{equation}
in our case we obtain
\begin{equation}
\rho^{\prime}+\rho\left(  1+\omega\right)  H+\frac{\Lambda^{\prime}c^{4}}{8\pi
G}+\rho\left(  \frac{G^{\prime}}{G}-4\frac{c^{\prime}}{c}\right)  =0,
\end{equation}
where $H=\frac{a^{\prime}}{a}+\frac{b^{\prime}}{b}+\frac{d^{\prime}}{d},$ but
if we take into account $\ $the condition, $T_{ij}^{;j}=0,$ it is obtained the
following set of equations:
\begin{align}
\rho^{\prime}+\rho\left(  1+\omega\right)  H  &  =0,\\
\frac{\Lambda^{\prime}c^{4}}{8\pi G\rho}+\frac{G^{\prime}}{G}-4\frac
{c^{\prime}}{c}  &  =0.
\end{align}

Therefore, the resulting FE\ are:\begin{widetext}
\begin{align}
\frac{a^{\prime}}{a}\frac{b^{\prime}}{b}+\frac{a^{\prime}}{a}\frac{d^{\prime}%
}{d}+\frac{d^{\prime}}{d}\frac{b^{\prime}}{b}-\frac{K^{2}}{4}\frac{a^{2}c^{2}%
}{b^{2}d^{2}}  &  =\frac{8\pi G}{c^{2}}\rho+\Lambda c^{2},\label{tam1}\\
\frac{b^{\prime\prime}}{b}+\frac{d^{\prime\prime}}{d}+\frac{d^{\prime}}%
{d}\frac{b^{\prime}}{b}-\frac{c^{\prime}}{c}\left(  \frac{b^{\prime}}{b}%
+\frac{d^{\prime}}{d}\right)  -\frac{3K^{2}}{4}\frac{a^{2}c^{2}}{b^{2}d^{2}}
&  =-\frac{8\pi G}{c^{2}}p+\Lambda c^{2},\label{tam2}\\
\frac{d^{\prime\prime}}{d}+\frac{a^{\prime}}{a}\frac{d^{\prime}}{d}%
+\frac{a^{\prime\prime}}{a}-\frac{c^{\prime}}{c}\left(  \frac{a^{\prime}}%
{a}+\frac{d^{\prime}}{d}\right)  +\frac{K^{2}}{4}\frac{a^{2}c^{2}}{b^{2}%
d^{2}}  &  =-\frac{8\pi G}{c^{2}}p+\Lambda c^{2},\label{tam3}\\
\frac{b^{\prime\prime}}{b}+\frac{a^{\prime}}{a}\frac{b^{\prime}}{b}%
+\frac{a^{\prime\prime}}{a}-\frac{c^{\prime}}{c}\left(  \frac{a^{\prime}}%
{a}+\frac{b^{\prime}}{b}\right)  +\frac{K^{2}}{4}\frac{a^{2}c^{2}}{b^{2}%
d^{2}}  &  =-\frac{8\pi G}{c^{2}}p+\Lambda c^{2},\label{tam4}\\
\rho^{\prime}+\rho\left(  1+\omega\right)  \left(  \frac{a^{\prime}}{a}%
+\frac{b^{\prime}}{b}+\frac{d^{\prime}}{d}\right)   &  =0.\label{tam5}\\
\frac{\Lambda^{\prime}c^{4}}{8\pi G\rho}+\frac{G^{\prime}}{G}-4\frac
{c^{\prime}}{c}  &  =0. \label{tam6}%
\end{align}
\end{widetext}

Note that we have taken into account the effects of a $c-$var into the
curvature tensor (see \cite{tony2} for a discussion about the issue).

In this model, the homothetic vector field is:%
\begin{align}
X  &  =\left(  \frac{\int cdt}{c(t)}\right)  \partial_{t}+\left(  1-\frac{\int
cdt}{c(t)}H_{1}\right)  x\partial_{x}+\nonumber\\
&  +\left(  1-\frac{\int cdt}{c(t)}H_{2}\right)  y\partial_{y}+\left(
1-\frac{\int cdt}{c(t)}H_{3}\right)  z\partial_{z},
\end{align}
with the following restrictions%
\begin{equation}
a=a_{0}\left(  \int cdt\right)  ^{\alpha_{1}},\,\,b=b_{0}\left(  \int
cdt\right)  ^{\alpha_{2}},\,\,d=d_{0}\left(  \int cdt\right)  ^{\alpha_{3}},
\label{scalesSS}%
\end{equation}
with $\left(  \alpha_{i}\right)  _{i=1}^{3}\in\mathbb{R},$ in such a way that
they must satisfy the relation $\alpha_{2}+\alpha_{3}-\alpha_{1}=1.$ Since we
expect to get a growing scale factors, we shall consider that $\left(  \int
cdt\right)  $ must be a positive growing time function. Note that $c$ only
needs to be integrable but it may be decreasing as we shall show below.

\begin{remark}
As it is observed, if $c=const.,$ we regain the usual homothetic vector field.
i.e.%
\begin{equation}
X=t\partial_{t}+\left(  1-tH_{1}\right)  x\partial_{x}+\left(  1-tH_{2}%
\right)  y\partial_{y}+\left(  1-tH_{3}\right)  z\partial_{z}, \label{homoc}%
\end{equation}
while the scale factors behave as%
\begin{equation}
a=a_{0}\left(  t\right)  ^{\alpha_{1}},\quad b=b_{0}\left(  t\right)
^{\alpha_{2}},\quad d=d_{0}\left(  t\right)  ^{\alpha_{3}},
\end{equation}
with $\alpha_{2}+\alpha_{3}-\alpha_{1}=1,$ as in the above studied cases.
\end{remark}

By defining%
\begin{equation}
H_{i}=\alpha_{i}\frac{c}{\int cdt}\qquad\Longrightarrow\qquad H=\alpha\frac
{c}{\int cdt}, \label{HSS}%
\end{equation}
where $\alpha=\sum_{i=1}^{3}\alpha_{i},$ we find, from eq. (\ref{tam5}), the
behavior of the energy density i.e.%
\begin{equation}
\rho=\rho_{0}\left(  \int cdt\right)  ^{-\gamma}, \label{rhoSS}%
\end{equation}
where $\gamma=(1+\omega)\alpha.$

In the same way it is easily calculated the shear%
\[
\sigma^{2}=\frac{1}{3c^{2}(t)}\left(  H_{1}^{2}+H_{2}^{2}+H_{3}^{2}-H_{1}%
H_{2}-H_{1}H_{3}-H_{2}H_{3}\right)  ,
\]
i.e.%
\begin{equation}
\sigma^{2}=\frac{1}{3}\left(  \sum_{i=1}^{3}\alpha_{i}^{2}-\sum_{i\neq
j}\alpha_{i}\alpha_{j}\right)  \left(  \int cdt\right)  ^{-2}.
\end{equation}
As it is observed all the quantities depend on $\int c(t)dt$. Now only rest to
calculate $G$ and $\Lambda$.

From eqs. (\ref{tam1} and \ref{rhoSS}) we get:%
\begin{equation}
A\left(  \frac{c}{\int c}\right)  ^{2}=\frac{8\pi G}{c^{2}}\rho_{0}\left(
\int c\right)  ^{-\gamma}+\Lambda c^{2},
\end{equation}
where we have written, for simplicity, $\int c$ instead of $\int cdt,$ and we
have set $A=\alpha_{1}\alpha_{2}+\alpha_{1}\alpha_{3}+\alpha_{2}\alpha
_{3}-\frac{K^{2}}{4},$ therefore%
\begin{equation}
\Lambda^{\prime}=-\frac{2Ac}{\left(  \int c\right)  ^{3}}-\frac{8\pi\rho_{0}%
G}{c^{4}\left(  \int c\right)  ^{\gamma}}\left[  \frac{G^{\prime}}{G}%
-4\frac{c^{\prime}}{c}-\gamma\frac{c}{\int c}\right]  .
\end{equation}

Now, taking into account eq. (\ref{tam6}), we get that%
\[
-\frac{c^{4}}{8\pi G\rho}\left[  \frac{2Ac}{\left(  \int c\right)  ^{3}}%
+\frac{8\pi\rho_{0}G}{c^{4}\left(  \int c\right)  ^{\gamma}}\left[
\frac{G^{\prime}}{G}-4\frac{c^{\prime}}{c}-\gamma\frac{c}{\int c}\right]
\right]  +
\]%
\begin{equation}
+\frac{G^{\prime}}{G}-4\frac{c^{\prime}}{c}=0,
\end{equation}
and hence we obtain%
\begin{equation}
G=\frac{A}{4\pi\rho_{0}\gamma}c^{4}\left(  \int c\right)  ^{\gamma-2},
\label{pili}%
\end{equation}
and in this way we find that the cosmological constant behaves as
\begin{equation}
\Lambda=A\left(  1-\frac{2}{\gamma}\right)  \left(  \int c\right)  ^{-2}.
\label{enma}%
\end{equation}

As we can see, from eqs. (\ref{pili} and \ref{enma}), we have that are
verified the following relationships%
\begin{equation}
\frac{G\rho}{c^{4}}\thickapprox\left(  \int c\right)  ^{-2},\qquad
\Lambda\left(  \int c\right)  ^{2}=const.
\end{equation}

Now, we will try to find the value of the constants $\left(  \alpha
_{i}\right)  _{i=1}^{3}.$ Taking into account the field eqs. (\ref{tam1}%
-\ref{tam4}) we find that, obviously eq. (\ref{tam1}) vanish, but from eqs.
(\ref{tam2}-\ref{tam4}) we get%
\begin{align}
a_{2}(a_{2}-1)+a_{3}\left(  a_{3}-1\right)  +a_{2}a_{3}-\frac{3K^{2}}{4}  &
=A\left(  \frac{\alpha-2}{\alpha}\right)  ,\label{nsis1}\\
a_{3}\left(  a_{3}-1\right)  +a_{1}a_{3}+a_{1}\left(  a_{1}-1\right)
+\frac{K^{2}}{4}  &  =A\left(  \frac{\alpha-2}{\alpha}\right)  ,\label{nsis2}%
\\
a_{2}(a_{2}-1)+a_{1}a_{2}+a_{1}(a_{1}-1)+\frac{K^{2}}{4}  &  =A\left(
\frac{\alpha-2}{\alpha}\right)  , \label{nsis3}%
\end{align}
finding therefore the same solution as in the above model, i.e.
\begin{equation}
a_{1}=2a_{2}-1,\quad a_{2}=a_{2},\quad K^{2}=-4a_{2}^{2}+6a_{2}-2,
\end{equation}
this solution has only sense if $a_{2}\in\left(  \frac{1}{2},1\right)  ,$ note
that for these values, $a_{1}>0.$

We would like to emphasize that to out line the system (\ref{nsis1}%
-\ref{nsis3}) it is essential to take into account the effects of a c-var into
the curvature tensor. If we do not consider such effects then the system
depends on $\int c$ and therefore it is not algebraic (see \cite{tony2} for a
detailed discussion).

Therefore the behaviour of the main quantities is the following one. The
energy density behaves as%
\begin{equation}
\rho=\rho_{0}\left(  \int c\left(  t\right)  dt\right)  ^{-\gamma},
\end{equation}
where $\gamma=\left(  \omega+1\right)  \alpha,$ and $\alpha=2\alpha_{2}%
+\alpha_{1},$ while Hubble and the deceleration parameters behave as%
\begin{equation}
H=\alpha\frac{c}{\int cdt},\quad q=\frac{1}{\alpha}\left(  1-\frac{c^{\prime}%
}{c}\frac{\int c}{c}\right)  -1,
\end{equation}
and the shear yields%
\[
\sigma^{2}=\frac{1}{3}\left(  \alpha_{2}-1\right)  ^{2}\left(  \int
cdt\right)  ^{-2}.
\]

On the other hand, $G$ behaves as, $G=G_{0}c^{4}\left(  \int cdt\right)
^{\gamma-2},$ where $G_{0}>0.$ Hence, its behaviour will depend on the
different values of $\gamma.$ If $\gamma\in\left(  0,2\right)  $ we find that
$G$ is a decreasing time function. If $\gamma=2$, $G=const.$ and all the model
collapses to the standard one studied in the above sections ($G=const.,$
$c=const.$ and $\Lambda=0$). While if $\gamma\in\left(  2,6\right)  $ then $G$
is a growing time function.

The cosmological constant behaves as%
\[
\Lambda=A\left(  1-\frac{2}{\gamma}\right)  \left(  \int cdt\right)
^{-2}=\Lambda_{0}\left(  \int cdt\right)  ^{-2},
\]
where the sign of $\Lambda_{0}$ depends on the different values of $\gamma,$%
\[
\Lambda_{0}=\left\{
\begin{array}
[c]{ll}%
<0 & if\quad\gamma\in\left(  0,2\right) \\
=0 & if\quad\gamma=2\\
>0 & if\quad\gamma\in\left(  2,6\right)
\end{array}
\right.  .
\]

Since our model is formally self-similar, then (\cite{Jantzen}-\cite{Wainwrit}%
) have shown, that all the quantities must follow a power law, so, we may
assume that \ for example, $c$ takes the following form: $c(t)=c_{0}%
t^{\epsilon},$ with $\epsilon\in\mathbb{R}.$ Hence, the scale factors behaves
as:%
\begin{equation}
a=a_{0}t^{\alpha_{1}\left(  \epsilon+1\right)  },\quad b=b_{0}t^{\alpha
_{2}\left(  \epsilon+1\right)  }=d,
\end{equation}
in such a way that they behave as growing time function if $\epsilon\in\left(
-1,\infty\right)  ,$ $\alpha_{i}>0.$ This means that $c$ may be a decreasing
time function. The homothetic vector field collapses to this one%
\begin{equation}
V=\frac{t}{\epsilon+1}\partial_{t}+\left(  1-a_{2}\right)  \left(
2x\partial_{x}+y\partial_{y}+z\partial_{z}\right)
\end{equation}

The Hubble parameter behaves as: $H=(4\alpha_{2}-1)\left(  \epsilon+1\right)
t^{-1}=\tilde{\alpha}t^{-1},$ so we find again the restriction for $\epsilon,$
$\epsilon\in\left(  -1,\infty\right)  .$ In the same way, we find from eq.
$\Lambda=\Lambda_{0}\left(  \int c\right)  ^{-2},$ that, since $\Lambda$ must
be a decreasing time function, this is only possible if $\epsilon\in\left(
-1,\infty\right)  ,$ the special case, $\epsilon=-1,$ is forbidden, note that
$\int cdt=\frac{c_{0}}{\epsilon+1}t^{\epsilon+1}>0,\ \forall\epsilon\in\left(
-1,\infty\right)  .$

So, we find that%
\begin{equation}
\rho\thickapprox t^{-\gamma\left(  \epsilon+1\right)  },\,\,G\thickapprox
t^{\gamma\left(  \epsilon+1\right)  +2\epsilon-2},\,\,\Lambda\thickapprox
t^{-2\left(  \epsilon+1\right)  },
\end{equation}
with $\epsilon\in\left(  -1,\infty\right)  ,$ and $\gamma=\left(
\omega+1\right)  \alpha=\left(  \omega+1\right)  (4a_{2}-1).$ Therefore the
energy density is always a decreasing time function if $\omega\in(-1,1]$ and
$\epsilon\in\left(  -1,\infty\right)  .$ As above, the behaviour of $G$
depends on $\left(  a_{2},\omega,\epsilon\right)  ,$ hence it may be a growing
function as well as a decreasing one. With regard to the cosmological
constant, the most important thing is to know its sign. Recalculating all the
computations we find that $\Lambda_{0}=\tilde{A}c_{0}^{-2}\left(  1-\frac
{2}{\tilde{\gamma}}\right)  $, where $\tilde{\gamma}=\gamma\left(
\epsilon+1\right)  =\left(  \omega+1\right)  (4a_{2}-1)\left(  \epsilon
+1\right)  $, in such a way that this sign is affected by the perturbing
parameter $\left(  \epsilon+1\right)  .$ We have set \ $\tilde{A}=\left(
5\alpha_{2}^{2}-2\alpha_{2}\right)  \left(  \epsilon+1\right)  ^{2}%
-\frac{K^{2}}{4},$ remember that, $\alpha_{1}=2\alpha_{2}-1.$

To clarify this results we may fix $\omega=1,$ $a_{2}=2/3$ and $\epsilon
=\pm1/2,$ finding%
\[%
\begin{array}
[c]{|c|c|c|c|}\hline
\epsilon & \rho & G & \Lambda_{0}\\\hline\hline
-1/2 & t^{-5/3} & t^{-4/3} & <0\\\hline
1/2 & t^{-5} & t^{4} & >0\\\hline
\end{array}
.
\]
Hence, if we take into account the current observation, then we must rule out
those values which make $\Lambda_{0}<0.$ Therefore we conclude that $c$ and
$G$ are growing time functions while $\Lambda$ is a positive time decreasing
time function.

With regard to the deceleration parameter we find that it has the following
behavior:
\begin{equation}
q=-1+\frac{1}{\tilde{\alpha}},
\end{equation}
where $\tilde{\alpha}=(4a_{2}-1)\left(  \epsilon+1\right)  ,$ indicating as
that it is quite unlikely that $\epsilon\rightarrow-1.$ We find that $q<0,$ if
$\tilde{\alpha}>1.$

\section{Conclusions.}

We have studied several perfect fluid Bianchi II models under the
self-similarity hypothesis. We have started our study reviewing the classical
solution and showing that it belongs to the LRS\ BII\ kind by calculating the
fourth Killing vector. The self-similar solution brings us to get a power law
solution for the scale factors i.e. they behave as $a=a_{0}t^{a_{1}},$
$b=b_{0}t^{a_{2}}=d,$ in such a way the the constants $a_{i}$ must satisfy the
relation $2a_{2}-a_{1}=1.$ The only drawback is that this solution is only
valid for an equation of state $\omega\in\left(  -1/3,1\right)  $. We have
also shown that this solution fits perfectly with the current observations for
$H$ and $q.$ With regard to its curvature behaviour we may conclude that the
model is singular. The Weyl parameter is quite small and the gravitational
entropy is constant as it is expected for a self-similar space-time. This
happens because the definition of gravitational entropy does not work in this
kind of space-times. We have finished our study of the classical model showing
that there is not solution for the vacuum model.

With regard to the perfect fluid model with $G$ and $\Lambda$ variable we have
arrived to the conclusion that its solution is quite similar to the obtained
one for the classical model. But we have been able to enlarge the set of
values for the equation of state, in this case is $\omega\in\left(
-1,\infty\right)  .$ As we have mentioned above the solution is similar to the
classical model so it also belongs to the LRS BII\ kind. With regard to the
behaviour of $G$ and $\Lambda$ we have showed that $G$ is a growing time
function while $\Lambda$ is a positive decreasing time function. We have ruled
out the rest of the solution taking into account the recent observation which
suggest us that $\Lambda_{0}>0$ and $q<0.$

In the VSL model we have arrived to similar conclusions as in the above model
i.e. the solution is valid for $\omega\in\left(  -1,\infty\right)  $. We have
started outlining the FE but taking into account the effects of a $c-$var into
the curvature tensor. In this approach is essential to take into account such
effects, in other way, it is impossible to get, after substitution, and
algebraic system of equations. We have also calculated the homothetic vector
field for the new metric as well as the constrains for the scale factors. In
this case all the quantities depend on $\int c(t)dt.$ By solving the
associated algebraic system we have obtained the same solution as in the above
studied case. Hence this solution belongs to the LRS\ BII type. In a generic
way we have discussed the behaviour of the main quantities ruling out
solutions like $\Lambda_{0}\leq0.$ If we assume a power law for $c(t)$,
$c=t^{\epsilon}$ say, then we arrive to the conclusion that it may be a
decreasing time function. The only restriction is $\int c(t)dt$ must be a
growing time function, hence $c$ must be integrable but it is allowed that it
may be decreasing. But if we consider the possibility of a $c$ decreasing then
$\Lambda_{0}<0.$ Therefore we have concluded that $G$ and $c$ are growing time
functions while $\Lambda$ is a positive decreasing time function.

\end{document}